\documentclass[preprint]{vgtc}             

\ifpdf
  \pdfoutput=1\relax                   
  \usepackage{graphicx}                
  \DeclareGraphicsExtensions{.pdf,.png,.jpg,.jpeg} 
\else
  \usepackage{graphicx}                
  \DeclareGraphicsExtensions{.eps}     
\fi%

\graphicspath{{figures/}{pictures/}{images/}{./}} 

\usepackage{microtype}                 
\PassOptionsToPackage{warn}{textcomp}  
\usepackage{textcomp}                  
\usepackage{mathptmx}                  
\usepackage{times}                     
\usepackage{cite}                      
\usepackage{tabu}                      
\usepackage{booktabs}                  
\usepackage[british]{babel}
\usepackage{csquotes}
\usepackage[flushleft]{threeparttable}
\usepackage{subfig}
\usepackage{float}
\usepackage{comment}
\usepackage{xcolor}

\onlineid{0}

\vgtccategory{Research}

\vgtcinsertpkg

\title{Influence of Personality and Communication Behavior of a Conversational Agent on User Experience and Social Presence in Augmented Reality}

\author{Katerina Koleva$^1$\thanks{e-mail: katerinagkoleva@gmail.com} 
\and Maurizio Vergari$^1$\thanks{e-mail: maurizio.vergari@tu-berlin.de} 
\and Tanja Kojić$^1$\thanks{e-mail: tanja.kojic@tu-berlin.de} 
\and Sebastian Möller$^{1,2}$\thanks{e-mail: sebastian.moeller@tu-berlin.de} 
\and Jan-Niklas Voigt-Antons$^3$\thanks{e-mail: jan-niklas.voigt-antons@hshl.de}}
\affiliation{\scriptsize $^1$Quality and Usability Lab, TU Berlin, $^2$German Research Center for Artificial Intelligence (DFKI)\\ $^3$Immersive Reality Lab, Hamm-Lippstadt University of Applied Sciences}

\abstract{A virtual embodiment can benefit conversational agents, but it is unclear how their personalities and non-verbal behavior influence the User Experience and Social Presence in Augmented Reality (AR). We asked 30 users to converse with a virtual assistant who gives recommendations about city activities. The participants interacted with two different personalities: Sammy, a cheerful blue mouse, and Olive, a serious green human-like agent. Each was presented with two body languages - happy/friendly and annoyed/unfriendly. We conclude how agent representation and humor affect User Experience aspects, and that body language is significant in the evaluation and perception of the AR agent.} 

\CCScatlist{
 \CCScatTwelve{Human-centered computing}{Human computer interaction (HCI)}{Interaction paradigms}{Mixed/augmented reality};
  \CCScatTwelve{Human-centered computing}{Interaction design}{}{};
}

\begin{document}


\firstsection{Introduction}

\maketitle

In the last decade, smart speech assistants have become more popular, while at the same time Augmented Reality (AR) applications have emerged as a powerful tool to help users in training, work, tourism, healthcare, or marketing. Mobile AR is beneficial for several reasons. First, it can be used on any mobile device, such as a smartphone or tablet, and second, it allows adding virtual objects to the real world, making it possible to use mobile AR applications everywhere, indoors or outdoors. Virtual Assistants (VAs) are digitally generated characters that provide information by voice, recognize the basic language, and can interact with the user by speech. Soon, speech assistants could be developed further, and we could have virtual assistants as part of each pair of AR glasses. To use all the benefits that visual embodiment offers, we should consider the different types of appearance and the body language of the VA. How we look and express ourselves verbally and non-verbally influences how we are perceived in our interactions with other humans daily. We expect that to apply also in the interaction between a user and a virtual assistant. Virtual characters who show emotions may increase the user's appreciation of a system. Showing emotion is closely related to body language, and the positive effect of polite gestures is already demonstrated in human-human communication \cite{bodyLan}. Such positive effects may also hold when communicating with a virtual assistant. This study can help better understand if the character and the non-verbal behavior of the speech assistant influence the User Experience, Social Presence, and overall User Perception in AR.

\section{Experiment Design}

\begin{figure}[h]
  \centering
\subfloat[]{
  \label{subfig: Sammy A}
    \includegraphics[width = 2cm, height =3cm]{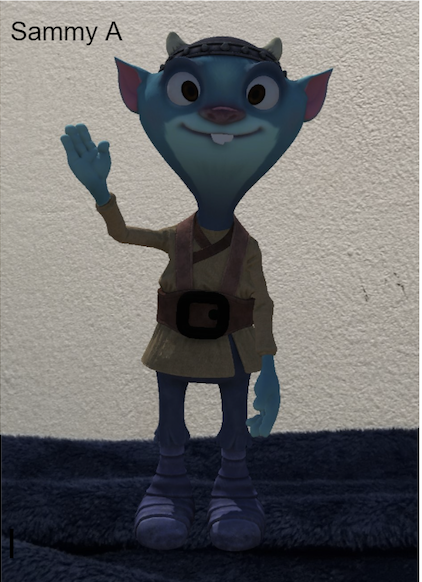}
    }
\subfloat[]{
  \label{subfig: Sammy B}
   \includegraphics[width = 2cm, height =3cm]{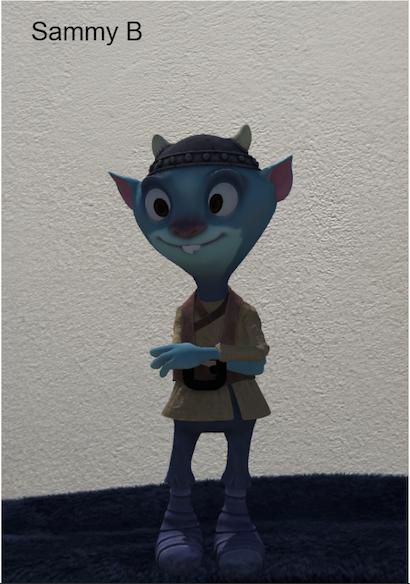}
    }
\subfloat[]
  {
  \label{subfig: Olive A}
   \includegraphics[width = 2cm, height =3cm]{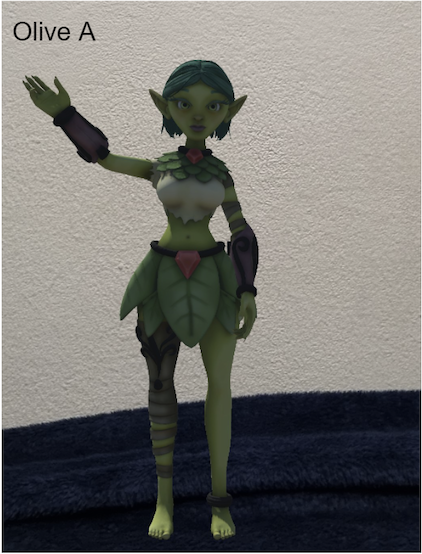}
  }
\subfloat[]
  {
  
  \label{subfig: Olive B}
   \includegraphics[width = 2cm, height =3cm]{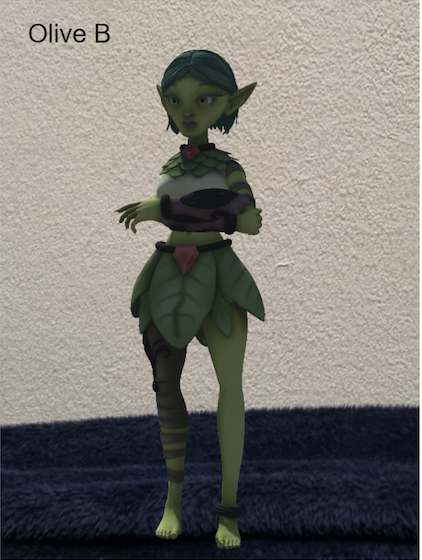}
  }
  \vspace*{-4px}
    \caption{The four versions of the virtual assistant.}
    \label{fig: Dynamic Game Tasks}
\end{figure}

It was decided to design two conversational agents who give recommendations about activities in the city with two different virtual bodies and speaking voices. Both of them were presented with two Communication Behaviors, expressed with different body gestures. Based on related work \cite{biocca03},\cite{mismatch}, which stated that users did not prefer the more anthropomorphic appearance of the agent, it was decided to develop Sammy as a non-human-like agent. An animated blue mouse character was chosen that has a big head and a cheerful facial expression (Fig. \ref{subfig: Sammy A}, \ref {subfig: Sammy B}). Sammy has a male voice and gives the impression of a virtual pet, a little helper, or a character from an animated kids' movie. Olive was created with the motivation to have a colder character with a less friendly personality. Some studies showed that a more anthropomorphic appearance would deliver a better experience to the users \cite{avatar-appearance}. Olive was designed to have a more anthropomorphic appearance, but still not entirely human-like (Fig. \ref{subfig: Olive A}, \ref {subfig: Olive B}). Non-behavioral cues for warmth and coldness of the agents indicated in the research of Cuddy et al. \cite{cuddy2008warmth} were considered when designing the two opposite communication behaviors. Also, non-verbal behavior in Embodied Conversational Agents has been shown to be really important for User Evaluation \cite{kramer}. The Communication Behavior, presented in the A version, was generally designed as more polite and friendlier to the user. Communication Behavior B was designed to be less friendly, and the agents were presented as disengaged and annoyed with the interaction.

\vspace{-7px}

 \vspace{-7px}
The user tests were designed as within-subject tests, which means each user interacted with the four different versions of the virtual assistant. The order in which the four assistants were used was randomized to prevent sequence effects 
After each condition, the participant was asked to fill out the User Experience Questionnaire plus (UEQ+) \cite{ueq+}, the Networked Minds Social Presence Measure for social presence \cite{socialpresence}, and some questions related to User Perception. At the end was asked \enquote{Did you think the body language of the virtual assistant matched their appearance and answers?} with \enquote{yes} or \enquote{no} answers, which was followed by the question \enquote{Why/Why not?}, requiring a short answer. 
The final questionnaire included some demographics, and some final questions.

\section{Results}

There were 30 participants aged between 19 and 35, from which 14 were female, and 16 were male, with an average age of 23.8 (SD = 2.94). They were students of different local universities, and most were familiar with some possible activities. This study used a two-way repeated measures ANOVA to determine if the different Personality and Communication Behavior influenced User Experience and Social Presence aspects in AR. An overview of all significant effects is given in Table \ref{tab:results}. 

\begin{table}[h!]
  \centering
   \caption{Two-way repeated-measures ANOVA significant results}
   
  \resizebox{\columnwidth}{!}{
\begin{tabular}{llcclrc}
\toprule
        Parameter & Effect & \( df_{n} \) & \( df_{d} \) & \( \hphantom{F}F \) & {\hspace{0.2cm}}\( p\hphantom{p.} \) & \(\eta_{G}^{2} \) \\
\midrule
        Personality & UX\_Attractiveness 
        & 1 & 29 & 18.905 & \textless.001 & .395 \\     
        Personality & UX\_Stimulation 
        & 1 & 29 & 8.719 & \textless.01 & .231  \\
        Personality & UX\_Novelty 
        & 1 & 29 & 5.078 & \textless.05 & .149 \\
        Personality & UX\_Usefulness 
        & 1 & 29 & 7.354 & \textless.05 & .202 \\
        Personality & UX\_Response.Behavior 
        & 1 & 29 & 13.918 & \textless.01 & .324 \\
        Personality & SP\_Co-presence 
        & 1 & 29 & 6.374 & \textless.05 & .180 \\
        Personality & SP\_Perceived.Emotional.Interdependence
        & 1 & 29 & 5.232 & \textless.05 & .153 \\
        Comm\_Behavior & UX\_Attractiveness 
        & 1 & 29 & 13.000 & \textless.001 & .310 \\
        Comm\_Behavior & UX\_Dependability 
        & 1 & 29 & 10.387 & \textless.01 & .264 \\
        Comm\_Behavior & UX\_Stimulation 
        & 1 & 29 & 6.707 & \textless.05 & .188 \\
        Comm\_Behavior & UX\_Novelty
        & 1 & 29 & 5.208 & \textless.05 & .152 \\
        Comm\_Behavior & UX\_Visual.Aesthetics
        & 1 & 29 & 7.785 & \textless.01 & .212 \\
        Comm\_Behavior & UX\_Response.Behavior
        & 1 & 29 & 10.240 & \textless.01 & .261 \\
        Comm\_Behavior & UX\_Response.Quality
        & 1 & 29 & 7.191 & \textless.05 & .199 \\
        Comm\_Behavior & SP\_Perceived.Message.Understanding 
        & 1 & 29 & 8.835 & \textless.01 & .234 \\
    
\bottomrule
    \end{tabular}%
    }
  \label{tab:results}%
  \vspace{0em}
\end{table}%

Participants rated the assistant named Sammy as more attractive (M = 5.733, SE = 0.155) than Olive (M = 5.033, SE = 0.187). Furthermore, Communication Behavior A (M = 5.829, SE = 0.152) was rated more attractive than Communication Behavior B (M = 4.938, SE = 0.231). In terms of dependability, communication behavior significantly influenced user perceptions. Communication Behavior A (M = 5.504, SE = 0.150) was rated more dependable than Communication Behavior B (M = 4.917, SE = 0.186). Participants found Sammy (M = 5.379, SE = 0.187) more stimulating than Olive (M = 4.963, SE = 0.170). Also, Communication Behavior A (M = 5.392, SE = 0.186) was more stimulating than Communication Behavior B (M = 4.950, SE = 0.184). Participants perceived Sammy (M = 5.442, SE = 0.211) as more novel than Olive (M = 5.125, SE = 0.237). Additionally, Communication Behavior A (M = 5.446, SE = 0.211) was considered more novel than Communication Behavior B (M = 5.121, SE = 0.238). Communication Behavior A (M = 5.592, SE = 0.191) was rated higher in visual aesthetics than Communication Behavior B (M = 5.092, SE = 0.231). Participants considered Sammy (M = 5.483, SE = 0.204) more useful than Olive (M = 5.092, SE = 0.223). Both personality and communication behavior significantly influenced the perceived response behavior of virtual assistants. Sammy (M = 5.113, SE = 0.196) was perceived as having better response behavior compared to Olive (M = 4.454, SE = 0.179). Furthermore, Communication Behavior A (M = 5.117, SE = 0.197) was rated more polite and trustworthy than Communication Behavior B (M = 4.450, SE = 0.195). Communication Behavior A (M = 5.633, SE = 0.204) was rated higher in response quality than Communication Behavior B (M = 5.342, SE = 0.225). When analysing Social Presence, users gave Sammy a higher co-presence score (M = 4.206, SE = 0.116) compared to Olive (M = 4.039, SE = 0.140). Communication Behavior A (M = 3.461, SE = 0.081) received a higher score in perceived message understanding than Communication Behavior B (M = 3.261, SE = 0.064). Sammy (M = 2.931, SE = 0.133) was rated higher than Olive (M = 2.744, SE = 0.129) in perceived emotional interdependence. In terms of user perception, most users (24 out of 30) preferred Sammy over Olive. The users liked Sammy's funnier and more likable appearance, his ability to make jokes, and his friendliness. Those who preferred Olive mentioned her beauty, sympathy, and confidence. When asked about their favorite version, 56.7\% chose Sammy A, 16.7\% chose Sammy B, and 13.3\% each chose Olive A and Olive B. Olive B was voted the least enjoyable by 70\% of users, while Olive A and Sammy B received 13.3\% of the votes each. Negative feedback for Olive B focused on her rudeness and unfriendly body language. Sammy A's body language was seen as consistent with his appearance and answers by 93.3\% of users, while Olive A received 70\% agreement. Sammy B and Olive B had lower percentages of agreement.

\section{Discussion \& Conclusion}

From the UX results we can argue that both the friendly appearance and the positive body language significantly affect how attractive, stimulating, and novel the virtual assistant is and how appropriate their response is. Non-verbal communication results are important because positive body language alone influences how predictable and appealing the users think the virtual assistant is, and showing that the personality does not have an effect on visual aesthetics, but the Communication Behavior does, and so the friendly, excited body language makes the assistant significantly more appealing to the user. It could be that for the context of CAs in AR for giving recommendations, nonverbal behavior includes a set of more sought-after traits. The results showed that a major effect played the Personality for the extent to which the user's emotional state is affected by that of the virtual assistant. Sammy, or the more friendly one, scored higher in perceived emotional interdependence. Furthermore, from the two different personalities again, Sammy evoked a significantly higher sense of co-presence. This could also be because a less anthropomorphic character represented him. Those results suggest that people feel more engaged and emotionally affected when talking to a character that seems from the imaginary world. For the overall User Perception, one of the most important conclusions is that the four versions of the virtual assistant were perceived differently, even though the base dialog system and given information were the same. The interaction with the assistants was described with words that people would use for human-to-human conversations. Moreover, although Sammy was chosen as more likable by 80\% of the people, some people still related to the more neutral anthropomorphic female assistant, which may be explained by the fact that people are different, and the users are looking for personalities in real life, as well as in the virtual world, that are suitable for them. It could be argued that the Non-verbal Communication Behavior of the assistants played an even bigger role in user evaluation than the Personality.

\subsection{Acknowledgements}
 This work was partly funded by the german Federal Ministry of Education and Research (BMBF) under the Software Campus program (grant number 01IS17052).


\bibliographystyle{abbrv-doi}

\bibliography{template}
\end{document}